\documentclass[aip, amsmath,amssymb, reprint]{revtex4-2}
\usepackage{graphicx}
\usepackage{amssymb}
\usepackage{epstopdf}
\usepackage{amsmath,dsfont, commath}
\usepackage{bm}
\usepackage{graphicx}
\usepackage{caption}
\usepackage{subfig}
\usepackage{overpic}
\usepackage{changes}
\usepackage{enumitem} 
\usepackage{subfig}
\usepackage{physics}
\usepackage{braket}
\usepackage{graphicx}
\usepackage{dcolumn}
\usepackage{bm}

\usepackage[utf8]{inputenc}
\usepackage[T1]{fontenc}
\usepackage{mathptmx}
\usepackage{etoolbox}

\newcommand{\RNum}[1]{\uppercase\expandafter{\romannumeral #1\relax}}
\newcommand{\be}{\begin{equation}}
	\newcommand{\ee}{\end{equation}}
\newcommand{\bea}{\begin{eqnarray}}
	\newcommand{\eea}{\end{eqnarray}}

\newcommand{\ba}{\begin{array}}
	\newcommand{\ea}{\end{array}}

\newcommand{\se}{Schr\"{o}dinger equation}
\newcommand{\bl}{\begin{flalign}}
	\newcommand{\enl}{\end{flalign}}

\renewcommand{\sec}[1]{Sec. \ref{#1}}

\renewcommand{\bf}[1]{\mathbf{#1}}
\usepackage[capitalize]{cleveref}% load last

\begin{document}

	%opening
	\title{Exponential convergence of the local diabatic representation for nonadiabatic models}%coupled electron-nuclear motion}
\author{Mo Sha}
\author{Bing Gu}
\email{gubing@westlake.edu.cn}
\affiliation{Department of Chemistry and Department of Physics, Westlake University, Hangzhou, Zhejiang, China, 310030}

\begin{abstract}
	The discrete variable local diabatic representation (LDR) provides a divergence-free framework for exact conical intersection dynamics simulation. In this work, we investigate the convergence with respect to the number of ``nuclear'' grid points and ``electronic'' states of LDR for the eigenvalue problems using coupled oscillator models. The performance of LDR is compared with traditional approaches based on the Born-Huang ansatz and on the crude adiabatic representation. Our results demonstrate that for weak vibronic couplings, LDR shows similar convergence rate as the exact Born-Huang representation including not only the first-order derivative couplings but also the diagonal Born-Oppenheimer  corrections and second-order derivative couplings. Surprisingly, for strong vibronic couplings, LDR shows a significant faster convergence rate with respect to the number of grid points, hence the number of electronic structure computations, than the exact Born-Huang representation. The crude adiabatic representation in generally shows a much slower convergence rate for all cases. The diagonal Born-Oppenheimer  corrections and second-order derivative couplings are found to be important in the Born-Huang framework. 

	%t provides equivalent accuracy while exhibiting a faster convergence rate in certain parameter regimes.
\end{abstract}
\maketitle

\section{Introduction}

% CIs are critical points where two or more adiabatic potential energy surfaces (PES) intersect, mediating ultrafast electronic transitions \cite{KEEFER2020Visualizing, KEEFER2020Visualizinga} and non-radiative relaxation \cite{HE2020Enantio, PRLJ2016How}. They play a critical role in photochemistry, photophysics, and biology\cite{CASIDA2012NonBornOppenheimer, Conical, GU2020Manipulatingc, GU2023Cavitya, Isomerization, MARQUES2012Fundamentals, MEAD1992geometric, NEVILLE2018Ultrafast, TRUHLAR1984Variational, XIE2016Nonadiabatica, XIE2017Constructive}. The complex and exceedingly fast nature of these dynamics makes their direct experimental characterization challenging, making accurate and efficient theoretical methods to simulate the dynamics near CIs essential for achieving a complete understanding of these fundamental chemical processes. 

The discrete variable local diabatic representation (LDR) is a numerically exact and divergence-free method for exact modeling of nonadiabatic quantum molecular dynamics, specifically for dynamics through conical intersections (CIs) \cite{GU2023DiscreteVariable, GU2024Nonadiabatic, XIE2025Topological, XIE2025Pathordered} whereby the non-Born-Oppenheimer effects become significant and must be taken into account in the nuclear motion \cite{DOMCKE2011Conical, WORTH2004BORNOPPENHEIMER, NEVILLE2018Ultrafast}. Such effects include nonadiabatic transitions, the diagonal Born-Oppenheimer corrections (DBOC), and geometric phase effects\cite{HAN2023Quantum}.
% Such effects include the nonadiabatic transitions, the diagonal Born–Oppenheimer correction (DBOC), and geometric phase effects. However, their treatment within the Born–Huang ansatz is problematic: first-order derivative (nonadiabatic) couplings diverge as the adiabatic gap closes, and the DBOC can exhibit a non-integrable singularity at conical intersections; the geometric phase, meanwhile, is a global topological effect (Berry phase) that cannot be represented as a simple local scalar correction and must be included via a vector potential\cite{HERZBERG1963Intersection, KOLOS1963Nonadiabatic, MEEK2016Wave, MEAD1979determination}. These failures of the local adiabatic expansion motivate alternative representations—such as the discrete variable local diabatic representation (LDR)—which recover non-Born-Oppenheimer physics from bounded global overlap quantities rather than singular derivative couplings.
The nonadiabatic transitions are a fundamental mechanism describing a wide range of phenomena in photochemistry and photophysics such as internal conversion, inter-system crossing, and photochemical reactions \cite{domcke2011, mead1979, mead1992}.
These transitions often occurs near conical intersections in polyatomic molecules or avoided crossings in diatomic molecules  where the energy gap between adiabatic potential energy surfaces becomes small, or even vanishes \cite{BIRCHER2018Nonadiabatic, WU2025Nonadiabatic, WANG2018Dynamical}.
In the traditional Born-Huang ansatz, these transitions are described by first- and second-order derivative couplings, but they diverge at conical intersections because the magnitude of the first-order derivative coupling is inversely proportional to the energy gap.

The diagonal Born-Oppenheimer corrections (DBOC),  the diagonal terms of the second-order derivative couplings, modify the adiabatic potential energy surfaces\cite{LENGSFIELD1986evaluation,YARKONY2019Diabatic}.  It has been found to be important for geometry optimization and harmonic frequencies in quantum chemistry simulations 
% and characterizing nonadiabatic tunneling in quantum dynamic simulation
\cite{HANDY1996adiabatic, VALEEV2003diagonal, XIE2017Dynamic, IOANNOU1996diagonal, SCHNEIDER2019Diagonal, XIE2017Constructive}. However, the singularity associated with the DBOC is particularly problematic: unlike the integrable singularities of first-derivative couplings, it is non-integrable and leads to undefined matrix elements unless the wave function rapidly decays off in the vicinity of the conical intersection. \cite{MEEK2016Wave, FEDOROV2019discontinuous, MEAD1983Electronic}.

Another topological feature of CIs is the geometric phase\cite{WITTIG2012Geometric, BERRY1998Paraxial, REQUIST2016Molecular}. A nuclear trajectory encircling a CI in configuration space acquires a geometric phase of $\pi$, which corresponds to a sign change of the adiabatic electronic wavefunction; to preserve single-valuedness of the total molecular wavefunction.  
% the compensating phase must reside in the nuclear component. 
 The geometric phase effect can strongly modify the nuclear wave functions by quantum interference in both nonadiabatic and adiabatic dynamics, and neglect of the geometric phase may even leads to qualitatively incorrect predictions for processes sensitive to nuclear wavefunction coherence near or around CIs\cite{RYABINKIN2013Geometric,RYABINKIN2017Geometric, WANG2024Impact, XIE2017Constructive}. 
 Incorporating the geometric phase into the nuclear motion is challenging, especially for trajectory-based semiclasssical methods. It can be included in the nuclear motion by a Mead-Truhlar vector potential \cite{MEAD1979determination, MEAD1983Erratum}, however constructing the vector potential from ab initio data is not straightforward \cite{malbon2016}.
% The geometric phase effect is a global topological effect (Berry phase) that cannot be represented as a simple local scalar correction and must be included via a vector potential.

In contrast to the Born-Huang framework, LDR describes \emph{all} non-Born-Oppenheimer effects, including nonadiabatic transitions, geometric phases, and DBOC, through the \emph{global} electronic overlap matrix, i.e., overlap between adiabatic many-electron wavefunctions at different nuclear geometries. One of its practical  advantages over the traditional Born-Huang framework is that the elements of the overlap matrix are inherently bounded within the range of $[-1, 1]$, which avoids the use of singular derivative couplings, allowing for robust calculations even near CIs. Furthermore, the construction of the overlap matrix does not require the adiabatic electronic states to be smooth with respect to the nuclear coordinates—a necessary gauge fixing condition for the definition of nonadiabatic couplings. The LDR method has been successfully applied to various vibronic coupling model Hamiltonians and for \textit{ab initio} calculations, demonstrating its accuracy and robustness\cite{GU2023DiscreteVariable, GU2024Nonadiabatic, ZHU2024Making} for conical intersection dynamics.

Nevertheless, a systematic benchmark of the numerical convergence of LDR is lacking. In this work, as a more stringent test, we study the numerical convergence of LDR for eigenvalue problems of nonadiabatic coupled oscillator models, whereby a high-frequency mode plays the role of ``electrons'' and a low-frequency mode plays the role of ``nuclei''. This model is chosen such that the ``electronic'' states can be analytically calculated, thus, the error of electronic structure computation is removed. Moreover, there is no conical intersection in such models, so that the conventional methods based on the Born-Huang ansatz can be applied without any singularity. This implies that there is no geometric phase effect meaning that a gauge fixing can be found such that the gauge connection vanishes with smooth real-valued electronic wavefunctions.  Thus, we do not need to construct the vector potential in the Born-Huang approach.  Such a gauge fixing is not required in LDR.  

We first consider an analytically solvable bilinearly coupled harmonic oscillator model and show that LDR converges exponentially with respect to both the number of nuclear grid points and the number of electronic states.
For the nonlinearly coupled oscillator models, we provide a comprehensive comparison among methods based on the LDR, conventional nonadiabatic methods based on the Born-Huang ansatz, and the crude adiabatic representation. 
Our results show that LDR method, even with an approximate overlap matrix from the linked product approximation\cite{XIE2025Pathordered}, exhibits the highest accuracy and fastest convergence rate among the tested methods for all models and in all parameter regimes. In the complete nuclear basis set limit, both LDR and the exact Born-Huang representation converge to the exact ground state energy with a relative error of $10^{-12} \sim 10^{-14}$. 
LDR methods converge much faster than the Born-Huang approach when the nonadiabatic effects become stronger.  The Born-Huang approach is found to be a poor approximation almost for all simulations and shows a slow convergence rate with respect to both the number of nuclear grid points and the number of electronic states. 
For the Born-Huang approach, all the non-Born-Oppenheimer corrections (including the first- and second-order couplings and DBOC) are shown to be important even for the ground state energy.
 Particularly, the second-order derivative coupling and DBOC, which are often neglected in quantum dynamics simulations, are shown to be crucial for the Born-Huang approach to achieve the exact limit. With only the first-order nonadiabatic coupling, the relative error for the ground and low-lying excited state energies is typically of the order of $10^{-3}$.
Despite the fact that the models do not exhibit large amplitude motion, the crude adiabatic representation is found to be a poor approximation almost for all simulations and shows a slow convergence rate with respect to both the number of nuclear grid points and the number of electronic states. 
 %and it is unable to achieve the exact limit of the ground state energy even with the complete nuclear basis set. Compared to the traditional Born-Huang ansatz, it delivers equivalent precision while demonstrating faster convergence in some instances. This analysis across a wide range of frequency ratios and coupling strengths confirms that LDR framework is a robust, accurate, and efficient tool for systems with challenging nonadiabatic effects.% We quantify performance by the relative error against exact solutions.%and two nonlinear models featuring different, more complex forms of electron-nuclear coupling. 

The remaining of the article is structured as follows. In \sec{sec:theory}, we briefly review the theories underlying the Born-Huang approach, crude adiabatic representation, and LDR. In \sec{sec:results}, we first benchmark LDR for the linear coupled harmonic oscillator model and then present the comparison among LDR, conventional methods based on the Born-Huang approach and the crude adiabatic representation for two nonlinearly coupled oscillator models. 
% Then, we present the results of our convergence analysis for each of the three coupled oscillator models—one linear and two nonlinear. 
\sec{sec:conclusion} summarizes.

Atomic units $\hbar=e=m_\text{e}=1$ are used throughout.

\section{Theory}\label{sec:theory}

\subsection{Born–Huang Representation}\label{BHF}

In the traditional Born–Huang approach, the total (time-independent) molecular wavefunction is expanded as\cite{BORN1996Dynamical}
\begin{equation}
\Psi(\mathbf{r},\mathbf{R}) = \sum _{\alpha=1}^N {\phi_\alpha(\mathbf{r};\mathbf{R})\chi_\alpha(\mathbf{R})} \label{BHE}
\end{equation}
where $\bf r$ ($\bf R$) refers to the electronic (nuclear) coordinates, $\chi_\alpha(\mathbf{R})$ is the nuclear wavefunction associated with the $\alpha$-th adiabatic electronic state $\phi_\alpha(\mathbf{r};\mathbf{R})$, which depends parametrically on the nuclear coordinates $\mathbf{R}$
\begin{equation}
\hat{H}_{\text{BO}}(\bf r; \mathbf{R})\ket{\phi_\alpha(\mathbf{R})} = V_\alpha(\mathbf{R})\ket{\phi_\alpha(\mathbf{R})}.
\end{equation}
Here, $\hat{H}_{\text{BO}}(\mathbf{R})$ is the electronic Born-Oppenheimer Hamiltonian, defined as the total Hamiltonian subtracting the nuclear kinetic energy operator, $\hat{H}_{\text{BO}}(\mathbf{R}) = \hat{H} - \hat{T}_\text{N}$, $V_\alpha(\mathbf{R})$ is the $\alpha$-th adiabatic potential energy surface (APES).

Inserting the Born-Huang expansion \cref{BHE} into the time-independent Schrödinger equation for the molecule $ \hat{H} \Psi(\mathbf{r},\mathbf{R}) = E \Psi(\mathbf{r},\mathbf{R}) $ , left-multiplying by $\phi_\beta(\mathbf{r};\mathbf{R})$, and integrating over the electronic degrees of freedom yields the nuclear \se \cite{BAER2006BornOppenheimer}
\begin{widetext}
\begin{equation}\label{BH-T}
\left( - \sum_{\mu} \frac{1}{2M_{\mu}} \nabla_{\mu}^2 + V_{\beta}(\mathbf{R}) \right) \chi_{\beta}(\mathbf{R}) + \sum_{\alpha} \left[ \sum_{\mu} -\frac{1}{2M_{\mu}}\qty(2 {F}_{\beta\alpha}^{\mu}(\mathbf{R})  \nabla_{\mu} + G_{\beta\alpha}^{\mu}(\mathbf{R}) )\right] \chi_{\alpha}(\mathbf{R}) = E \chi_{\beta}(\mathbf{R}) 
\end{equation}
\end{widetext}
where $\nabla_\mu = \pd{}{R_\mu}$, represents the partial derivative with respect to the nuclear coordinate $R_\mu$ with mass $M_\mu$. 
Here 
\be {F^{\mu}_{\beta\alpha}}(\mathbf{R})=\langle\phi_{\beta}(\mathbf{R})|\nabla_{\mu}|\phi_{\alpha}(\mathbf{R})\rangle_{\mathbf{r}} 
\ee 
 is the first-order derivative coupling (NAC),  ${\langle \cdots \rangle}_\mathbf{r}$ denotes integration over electronic degrees of freedom, and
 \be G^\mu_{\beta\alpha}(\mathbf{R})=\langle\phi_{\beta}(\mathbf{R})|\nabla^2_{\mu}|\phi_{\alpha}(\mathbf{R})\rangle_{\mathbf{r}} 
 \ee denotes the second-order derivative coupling (SDC) for $\beta \ne \alpha$ and the DBOC for $\beta = \alpha$ that modifies the adiabatic potential energy surfaces. The gauge connections are chosen to vanish $\bf F_{\alpha \alpha} = 0$, so-called parallel transport gauge. Note that this gauge fixing is impossible in the presence of a conical intersection. 
Making use of the identity $\sum_{\gamma=0}^\infty|\phi_{\gamma}(\mathbf{R})\rangle\langle\phi_{\gamma}(\mathbf{R})|=\hat{I}$, SDC can also be expressed as
\begin{equation}\label{sdc}
G^{\mu}_{\beta\alpha} = \sum_{\gamma=0}^\infty {F}^{\mu}_{\beta\gamma} {F}^{\mu}_{\gamma\alpha} + \nabla_{\mu} {F}^{\mu}_{\beta\alpha}
\end{equation}
where $\gamma$ runs over all electronic states. 
% The diagonal terms of the second-order derivative coupling, $G^{\mu}_{\alpha\alpha}(\mathbf{R})=\langle\phi_{\alpha}(\mathbf{R})|\nabla^2_{\mu}|\phi_{\alpha}(\mathbf{R})\rangle_{\mathbf{r}}$, are known as the Diagonal Born-Oppenheimer Correction (DBOC) to the adiabatic potential energy surface. 
The NAC and SDC account for nonadiabatic transitions so that if both the first- and second-order derivative couplings are neglected such that the molecule is not allowed to make nonadiabatic transitions,  the adiabatic limit is recovered. If the DBOC and geometric phase effects are further neglected, it yields the Born-Oppenheimer approximation.

\subsection{Crude Adiabatic Representation}

The crude adiabatic representation is in fact the simplest form of a diabatic representation in the sense that the electronic states do not vary with nuclear geometries \cite{TANNOR2007Introduction}. 
Instead of the adiabatic electronic states that depend on the nuclear configuration, it employs the adiabatic electronic states at a fixed reference geometry $\ket{\phi_{\alpha}(\bf R_0)}$. 
%that are independent of the nuclear geometry $\mathbf{R}$, evaluated at a fixed reference geometry $\mathbf{R_0}$, i.e., $\phi_{\alpha}(\mathbf{r}; \mathbf{R}) \to \phi_{\alpha}(\mathbf{r}; \mathbf{R_0})$. 
The total vibronic wavefunction is then written as 
\begin{equation}
\Psi(\mathbf{r},\mathbf{R}) = \sum_{\alpha = 0}^{N-1} \phi_{\alpha}(\mathbf{r}; \mathbf{R_0}) \chi_\alpha(\mathbf{R})
\label{CAR}
\end{equation}
If we define a potential difference operator be $\Delta \hat{H}(\mathbf{R}) = \hat{H}_{\mathrm{BO}}(\mathbf{R}) - \hat{H}_{\mathrm{BO}}(\mathbf{R_0})$, which represents the change in the electronic potential as the nuclei move away from the reference geometry $\mathbf{R_0}$. The total Hamiltonian can be rewritten as $\hat{H} = \hat{T}_\text{N} + \hat{H}_{\mathrm{BO}}(\mathbf{R_0}) + \Delta \hat{H}(\mathbf{R})$.
Substituting Eq.~\eqref{CAR} into the time-independent Schrödinger equation, left-multiplying by $\phi_\beta(\mathbf{r};\mathbf{R_0})$, and integrating over the electronic degrees of freedom yields

\begin{multline}
    \left( - \sum_{\mu} \frac{1}{2M_{\mu}} \nabla_{\mu}^2 + V_{\beta}(\mathbf{R_0})\right) \chi_{\beta}(\mathbf{R}) \\+ \sum_{\alpha} \Delta H_{\beta\alpha}(\mathbf{R}) \chi_{\alpha}(\mathbf{R}) = E \chi_{\beta}(\mathbf{R})
\end{multline}
where
$
\Delta H_{\beta\alpha}(\mathbf R)
=\bigl\langle\phi_\beta(\mathbf r;\mathbf R_0)\bigm|
\Delta \hat{H}(\mathbf R)
\bigm|\phi_\alpha(\mathbf r;\mathbf R_0)\bigr\rangle_{\!\mathbf r} 
$
is the diabatic potential energy matrix\cite{MASKRI2022moving}.

The advantage of this representation is that by employing an electronic basis $\ket{\phi_{\alpha}(\bf R_0)}$ fixed at a reference geometry, it entirely circumvents the calculation of derivative couplings. This avoids the singularity problem associated with derivative couplings near conical intersections.
However, this representation has a significant limitation that prevents it from being used in molecular processes involving large amplitudes motion. Although the crude adiabatic representation is formally exact in the complete electronic state limit (involving infinitely many electronic states), a truncation of electronic states is inevitable for any process of chemical interest. Upon truncation, the crude adiabatic representation is poor at describing motion that deviates far from the reference geometry.

\subsection{Local Diabatic Representation}
Conceptually, the local diabatic representation can be considered as a local generalization of the crude adiabatic representation with many reference geometries instead of one.
%The Local Diabatic Representation (LDR) describes the nuclear motion using
A crucial point is that the reference geometries are chosen through a discrete variable representation (DVR) \cite{COLBERT1992novel, LIGHT1985Generalized} of the reactive coordinate operators.
Any DVR basis set can be employed within LDR, the optimal choice depends on the specific process and boundary conditions. This immediately opens up the powerful DVR toolbox that has been developed for wave packet dynamics to nonadiabatic quantum molecular dynamics, and in particular, for conical intersection dynamics, such as the non-direct-product DVR \cite{XIE2024NondirectProduct}.

In contrast to the Born-Huang expansion, LDR uses an ansatz where the adiabatic electronic states are evaluated only at a discrete set of nuclear geometries, $\mathbf{R_n}$, predetermined by the DVR grid
\begin{equation}
\Psi(\mathbf{r}, \mathbf{R},t) = \sum_{\mathbf{n}, \alpha} C_{\mathbf{n}\alpha} (t)\phi_\alpha(\mathbf{r}; \mathbf{R_n}) \chi_{\mathbf{n}}(\mathbf{R})
%\equiv \sum_{\bf{n}\alpha}{C_{\mathbf{n}\alpha}(t)\langle \mathbf{r}; \mathbf{R}  |\mathbf{n}\alpha \rangle}
\label{eq:ansatz}
\end{equation}
Here, the vibronic basis functions
%$\{ \langle \mathbf{r}, \mathbf{R} | \mathbf{n}\alpha \rangle = 
$\set{\phi_\alpha(\mathbf{r}; \mathbf{R_n}) \chi_{\mathbf{n}}(\mathbf{R}) }$ are a direct product of the nuclear DVR basis functions, $\chi_{\mathbf{n}}(\mathbf{R})$, a localized nuclear basis function centered at the geometry $\mathbf{R_n}$, and the adiabatic electronic states, $\phi_\alpha(\mathbf{r}; \mathbf{R_n})$ are eigenstates of the electronic Hamiltonian at this reference geometry $\hat{H}_\text{BO}(\bf r; \bf R_{\bf n})$, $
\hat{H}_\text{BO}(\bf r; \bf R_{\bf n}){\phi_\alpha(\mathbf{r}; \mathbf{R_n})} = V_\alpha(\mathbf{R_n}){\phi_\alpha(\mathbf{r}; \mathbf{R_n})}
\label{eq:bo_eigenproblem}
$. 
For brevity, the composite basis state is denoted as $\ket{\mathbf{n}\alpha} \equiv \ket{\phi_\alpha(\mathbf{R_n})} \otimes \ket{\chi_{\mathbf{n}}}$, with $C_{\mathbf{n}\alpha}(t)$ being the time-dependent expansion coefficients.
For a multi-dimensional system, the DVR grid points (and thus the reference geometries $\mathbf{R_n}$) can be generated from a direct product of 1D grids. 

%These electronic states and their corresponding potential energies, $E_\alpha(\mathbf{R_n})$, are the solutions to the time-independent electronic Schrödinger equation for a fixed nuclear geometry $\mathbf{R_n}$:
% \begin{equation}
% \ket{\mathbf{R_n}} = \ket{R_{n_1}^1} \otimes \ket{R_{n_2}^2} \cdots \otimes \ket{R_{n_d}^d}
% \end{equation}
% where $\mathbf{n} = \{n_1, n_2, \dots, n_d\}$ is a multi-index.

LDR inherits the advantages of a DVR basis set in terms of the construction of Hamiltonian matrix elements. As the nuclear kinetic energy operator $\hat{T}_\text{N}$ acts solely on the nuclear space, its matrix elements in LDR basis are given by
\begin{align}
\braket{ \mathbf{m}\beta | \hat{T}_\text{N} | \mathbf{n}\alpha } = A_{\mathbf{m}\mathbf{n}}^{\beta\alpha} \braket{ \mathbf{m} | \hat{T}_\text{N} | \mathbf{n} }_{\mathbf{R}}
\end{align}
where $\braket{ \mathbf{m} | \hat{T}_\text{N} | \mathbf{n} }_{\mathbf{R}}$ is the kinetic energy matrix element in the nuclear DVR basis, which can often be analytically calculated and $A_{\mathbf{m}\mathbf{n}}^{\beta\alpha}$ is the electronic overlap matrix, defined as
\begin{equation}
  \label{ovlp}
A_{\mathbf{m}\mathbf{n}}^{\beta\alpha} = \braket{\phi_\beta(\mathbf{R_m}) | \phi_\alpha(\mathbf{R_n})}_\mathbf{r}
\end{equation}
Because the DVR basis state $\ket{\mathbf{n}}$ is an eigenstate of the position operator, the matrix elements of the electronic Hamiltonian are diagonal
\begin{equation}
\braket{ \mathbf{m}\beta | \hat{H}_{\text{BO}} | \mathbf{n}\alpha } = V_\alpha(\mathbf{R_n}) \delta_{\mathbf{m}\mathbf{n}} \delta_{\beta\alpha}
\end{equation}

In LDR, the electronic overlap matrix, $A_{\mathbf{m}\mathbf{n}}^{\beta\alpha}$, plays a fundamental role in describing non-adiabatic quantum molecular dynamics, as it encodes all effects beyond the Born-Oppenheimer approximation. 
This can be seen by noting that if the electronic state information is removed, i.e., assuming that the adiabatic states do not vary with nuclear geometries, $A_{\mathbf{m}, \mathbf{n} }^{\beta \alpha} = \delta_{\beta \alpha}$, LDR simply reduces to the Born-Oppenheimer limit.

As the overlap matrix elements are bounded in the range of $[-1,1]$, LDR avoids all the divergences in the Born-Huang representation and is numerically stable even near conical intersections.

\textit{Ab initio} simulation of the global overlap matrix can be computationally demanding. For a $d$-dimensional system, the computational cost scales as $\mathcal{O}(n^{2d})$, where $n$ is the number of grid points per degree of freedom.
The Linked Product Approximation (LPA) was recently introduced to reduce this complexity \cite{XIE2025Pathordered}.
This approximation constructs the global overlap matrix from local, nearest-neighbor overlaps (links), circumventing the need to compute overlaps for all pairs of nuclear configurations. Specifically, a long-range overlap matrix element is approximated by a product of links along a predefined path connecting the two geometries.
Given two distinct d-dimensional configurations labeled by $\mathbf{m}=(m_1,m_2,\ldots,m_d)$ and $\mathbf{n}=(n_1,n_2,\ldots,n_d)$, a nearest-neighbor path $\gamma$ can be given by the sequence $\gamma: \mathbf{n} \rightarrow (m_1,n_2 ,\ldots ,n_d)\rightarrow (m_1,m_2 , \ldots , n_d)\rightarrow\cdots \rightarrow \mathbf{m}$, the corresponding overlap matrix element is approximated by the ordered matrix product
\begin{equation}
	\mathbf{A}_{\mathbf{m}\mathbf{n}} \approx \prod_{k=0}^{L-1} \mathbf{A}_{\gamma_{k+1}, \gamma_{k}}
\label{eq:lpa}
\end{equation}
$L = \lVert \mathbf{n} - \mathbf{m} \rVert_1$ is the path length, and the path runs from $\gamma_0 = \mathbf{n}$ to $\gamma_{L} = \mathbf{m}$. This approximation becomes exact in the complete basis set limit.
While the LPA has been demonstrated to be highly accurate for conical intersection dynamics, we provide a more stringent assessment of this approximation on the stability and accuracy for eigenvalue problems.

% For instance, given two distant configurations labeled by $\mathbf{m}=(1,1)$ and $\mathbf{n}=(2,4)$, one possible path $\gamma$ that connect two configurations can be defined as:
% $
% \gamma: (1,1) \rightarrow (1,2) \rightarrow (1,3) \rightarrow (1,4) \rightarrow (2,4) 
% $.

\section{Models and Computational Results}
\label{sec:results}

%Electron-nuclear (vibronic) coupling is a crucial factor in nonadiabatic processes, particularly in the vicinity of conical intersections. To capture the essential features of these interactions, 
We consider coupled oscillator models where the ``electronic'' states can be analytically calculated. Specifically, the high-frequency mode $x$ plays the role of the ``electrons'', while the low-energy mode $y$ represents ``nuclei''. 
We first use a analytically solvable coupled harmonic oscillator model to benchmark the accuracy of LDR. 
We then extend the analysis to two nonlinearly coupled oscillator models comparing the convergence of LDR-based methods, the conventional Born-Huang-based methods, and the crude adiabatic representation.

%These models are designed to test the robustness of LDR method for capturing nonadiabatic effects in more challenging scenarios.

\subsection{Coupled Harmonic Oscillator Model}\label{sec:linear}

For a linearly coupled harmonic oscillator model, an analytical solution can be readily obtained. The Hamiltonian for such a system, denoted $\hat{H}_\text{\RNum{1}}$, is given by:
\begin{equation}
\hat{H}_\text{\RNum{1}} = \frac{1}{2}\omega_1 \qty( \hat{p}_x^2 + \hat{x}^2) + \frac{1}{2}\omega_y\qty( \hat{p}_y^2 + \hat{y}^2) + \frac{1}{2} g\hat{x}\hat{y}
\end{equation}
Here, $\hat{x}$ and $\hat{y}$ are the coordinate operators and $\hat{p}_x=-i\nabla_x$ and $\hat{p}_y=-i\nabla_y$ are the momentum operators. The coupling constant $g$ has unit of energy. 
Without loss of generality, we set $\omega_y = 1$, i.e., $\omega_y$ is the energy unit. 
We investigate three different ``electronic'' frequencies: $\omega_1 = 1.0, 3.0,$ and $10.0$. For each frequency, we consider two distinct coupling strengths, $g = 0.5$ and $0.80$, representing the weak and strong coupling regimes, respectively.

This model can be analytically solved by diagonalizing its Hessian matrix
\begin{equation}
\mathbf{H} =
\begin{pmatrix}
\omega_1^2 & \frac{g}{2}\sqrt{\omega_1} \\
\frac{g}{2}\sqrt{\omega_1} & 1
\end{pmatrix}
\end{equation}
The normal mode frequencies, $\Omega_1$ and $\Omega_2$, can be obtained from the eigenvalues of the matrix $\mathbf{H}$. The exact energy eigenvalues of the Hamiltonian $\hat{H}_\text{\RNum{1}}$ are therefore the sum of two independent quantum harmonic oscillators:
\begin{equation}
E_{n_1,n_2} = \Omega_1\left(n_1 + \frac{1}{2}\right) + \Omega_2\left(n_2 + \frac{1}{2}\right)
\end{equation}
where $n_1, n_2 = 0, 1, \cdots $ are the quantum numbers for the two normal modes, respectively.

Within the Born-Oppenheimer approximation, the ``electronic'' Hamiltonian for this model is
%is defined with $x$ as the electronic and $y$ as the nuclear coordinate:
\[
\hat{H}_{\text{BO}}(y) = \frac{1}{2}\omega_1 \hat{p}_x^2 + \frac{1}{2}\omega_1 \hat{x}^2 + \frac{1}{2}y^2 + \frac{1}{2} g\hat{x}y
\]
The eigenvalues of $\hat{H}_{\text{BO}}$ with respect to $x$ for fixed values of $y$ yield the APESs for nuclear motion along the $y$-coordinate.
The corresponding eigenstates of $\hat{H}_{\text{BO}}$, denoted as $\phi_{\alpha}(x;y)$, are functions of the electronic coordinate $x$ that depend parametrically on the nuclear coordinate $y$. They can be calculated analytically as the ``electronic'' Hamiltonian is a displaced harmonic oscillator
\begin{equation}
  \phi_{\alpha}(x;y) = \varphi_{\alpha}\bigl(x - \Delta(y)\bigr)
\end{equation}
where $\varphi_{\alpha}(x)$ are the eigenstates of $\frac{1}{2}\omega_1 \qty( \hat{p}_x^2 + \hat{x}^2) $.
Here, the displacement function $  \Delta(y) = -\frac{gy}{2\omega_1}$
depends on the nuclear coordinate $y$ and arises from the bilinear coupling term $g\hat{x}{y}$.

The electronic overlap matrix elements $A_{\beta \alpha}(y, y')$ between two different nuclear geometries, $y$ and $y'$, can be calculated analytically. The derivation starts with the ground electronic states ($\alpha=0, \beta=0$) and proceeds recursively for the excited states ($\alpha>0$ or $\beta>0$). The relations are
\begin{subequations}\label{eq:overlap_full}
\begin{align}
    % --- 公式 (a) ---
    A_{0,0}(y,y') =&\exp\!\Bigl[-\frac{\Delta _{y,y'}^2}{4}\Bigr] \label{eq:overlap_base} \\
    % --- 公式 (b) ---
    A_{\beta,\alpha}(y,y') 
        & =\tfrac{1}{2}\sqrt{\tfrac{2}{\beta}}\; \Delta_{y,y'}\; A_{\beta-1,\alpha}(y,y')\nonumber\\
        &\quad + \sqrt{\tfrac{\alpha}{\beta}}\; A_{\beta-1,\alpha-1}(y,y') \label{eq:overlap_rec1} \\
    % --- 公式 (c) 
        & =-\tfrac{1}{2}\sqrt{\tfrac{2}{\alpha}}\; \Delta_{y,y'}\; A_{\beta,\alpha-1}(y,y')\nonumber \\
        & \quad+ \sqrt{\tfrac{\beta}{\alpha}}\; A_{\beta-1,\alpha-1}(y,y') \label{eq:overlap_rec2}
\end{align}
\end{subequations}
Here, $\Delta_{y,y'} = \Delta(y) - \Delta(y')$, and the indices $\alpha, \beta \in \{0, 1, 2, \dots\}$ label the electronic states. The recursive calculation uses relation Eq.~\eqref{eq:overlap_rec1} for states where $\alpha=0$, and relation Eq.~\eqref{eq:overlap_rec2} for states where $\beta=0$. Elements with a negative index are zero.

\begin{figure*}[htbp]
\includegraphics[width=0.650\textwidth]{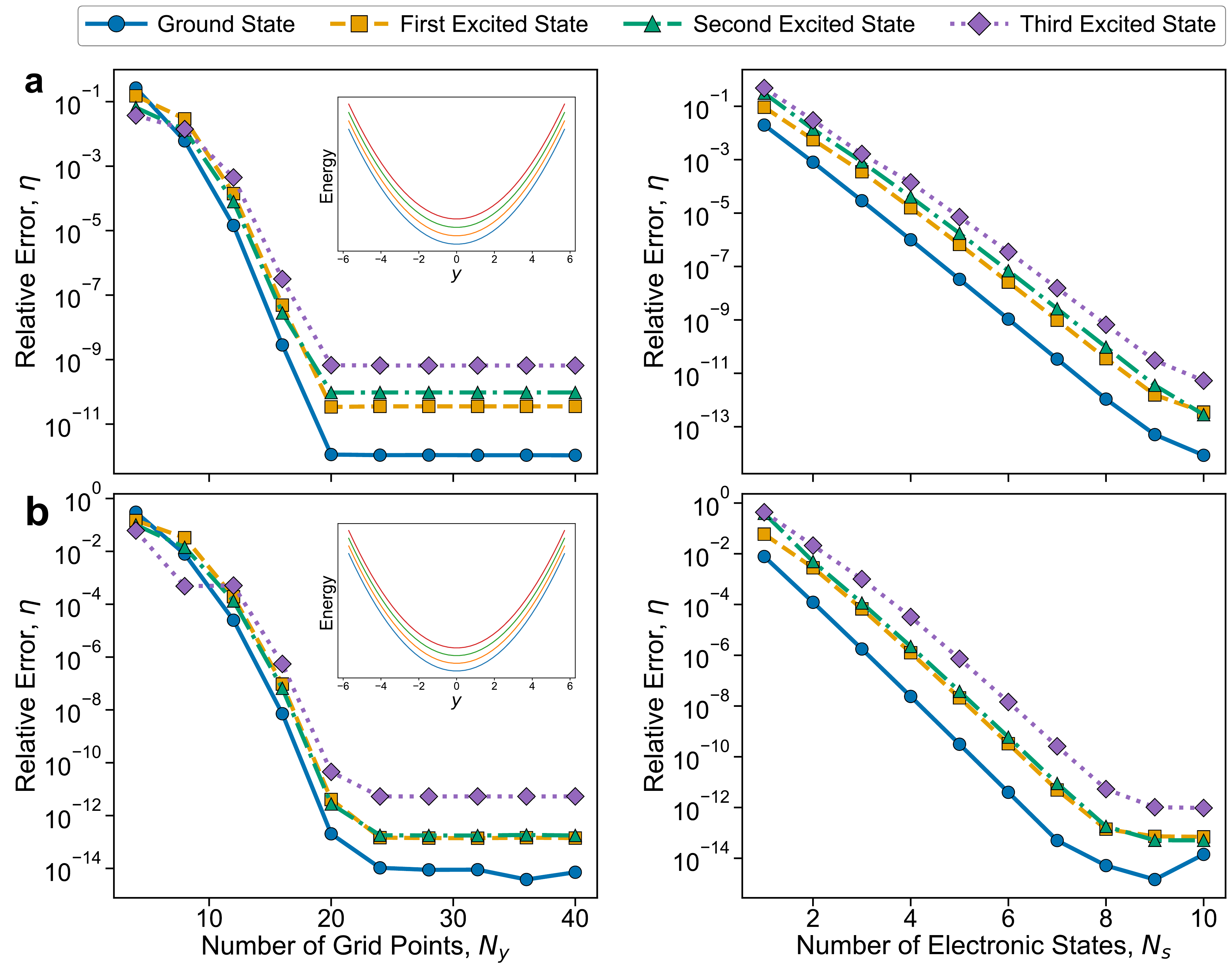} %插入图片，[]中设置图片大小，{}中是图片文件名
\caption{
Convergence tests at $\omega_1 = 1.0$ for two different coupling strengths.
(\textbf{a}) Strong coupling ($g=0.80$): An accuracy of $10^{-12}$ is achieved with 20 grid points, while $10^{-14}$ is achieved with a basis of 10 electronic states.
(\textbf{b}) Weak coupling ($g=0.5$): An accuracy of $10^{-14}$ is achieved with 20 grid points, while $10^{-14}$ is achieved with a basis of 8 electronic states.}\label{fig:analytical1}
\end{figure*}

\begin{figure*}[htbp]
\includegraphics[width=0.650\textwidth]{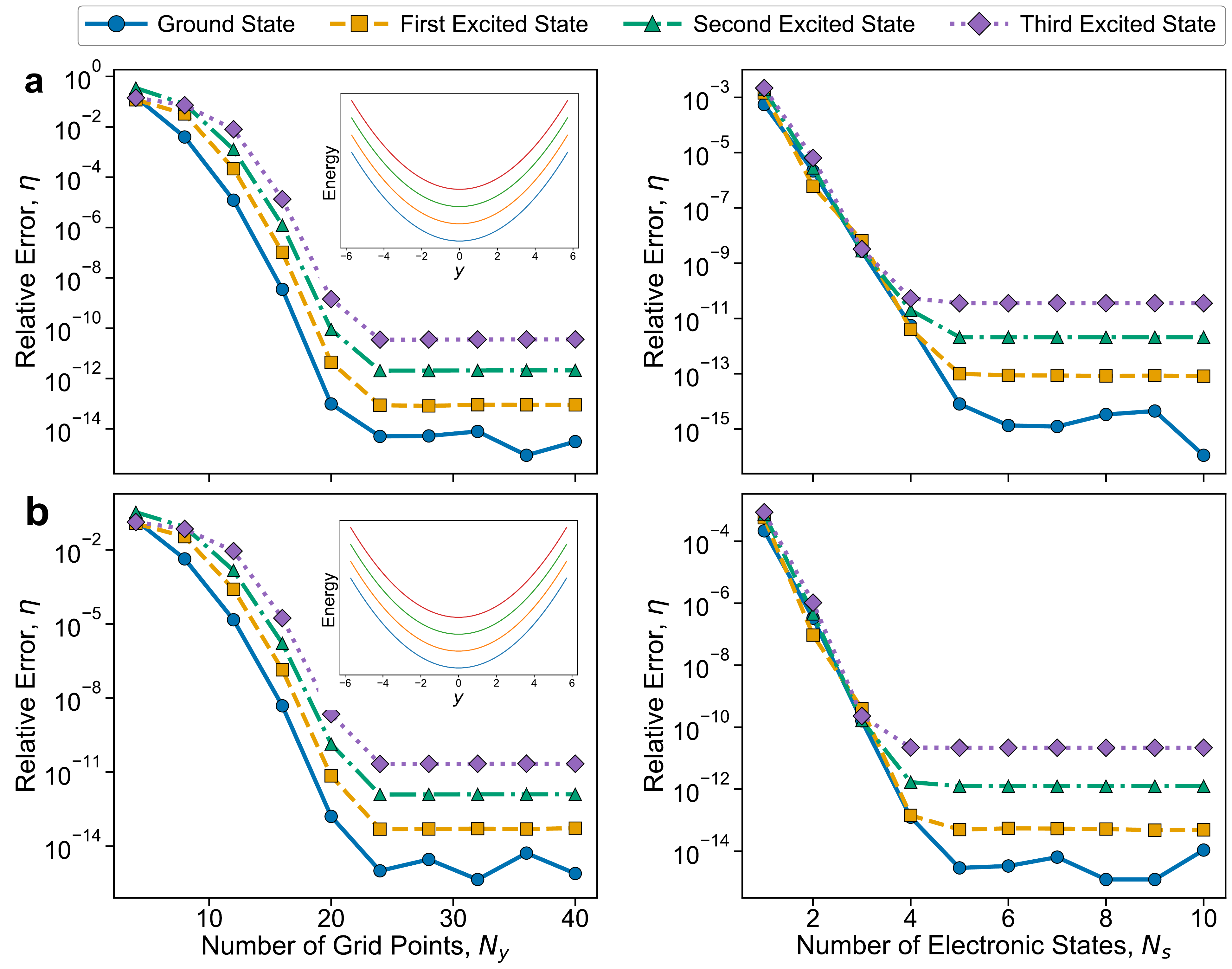} %插入图片，[]中设置图片大小，{}中是图片文件名 
  \caption{
  Convergence tests at $\omega_1 = 3.0$ for two different coupling strengths.
  (\textbf{a}) Strong coupling ($g=0.80$): An accuracy of $10^{-15}$ is achieved for the ground state with a basis of 6 electronic states.
  (\textbf{b}) Weak coupling ($g=0.5$): The same level of accuracy is reached using 5 electronic states.
  The convergence with respect to the nuclear grid is obtained with approximately 20 grid points to reach a relative error of $10^{-15}$ for both coupling strengths.
  }\label{fig:analytical2}
\end{figure*}

\begin{figure*}[htbp]
\includegraphics[width=0.650\textwidth]{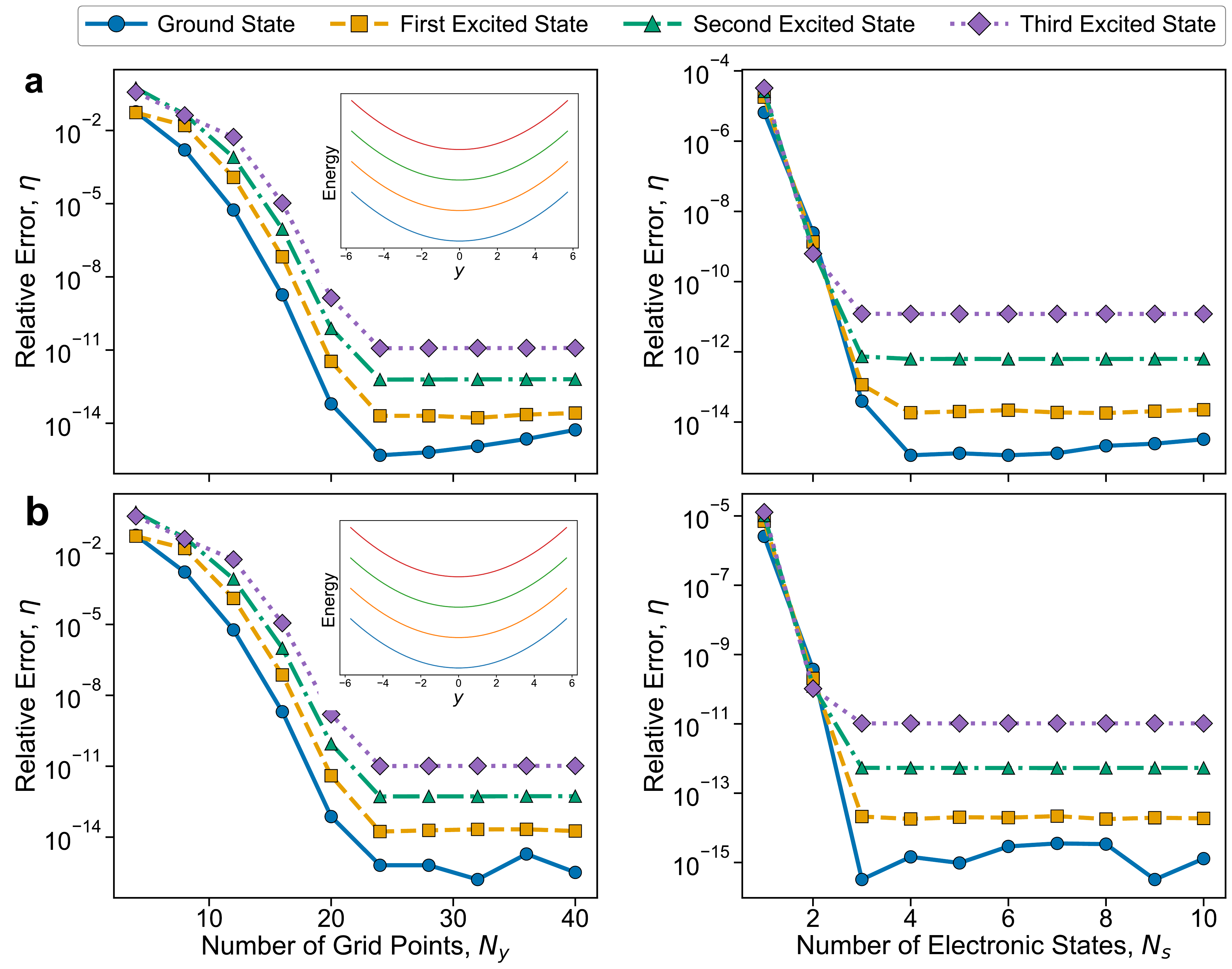}
\caption{
  Convergence tests at $\omega_1 = 10.0$ for two different coupling strengths.
  (\textbf{a}) Strong coupling ($g=0.80$): An accuracy of $10^{-15}$ is achieved with a basis of at least 4 electronic states.
  (\textbf{b}) Weak coupling ($g=0.5$): The same level of accuracy is reached using only 3 electronic states.
  The nuclear grid required approximately 20 points to reach a relative error of $10^{-15}$ for both coupling strengths. And the first excited state converged to an accuracy of $10^{-14}$.
}\label{fig:analytical3}
\end{figure*}

\begin{figure*}[htbp]
	\includegraphics[width=0.65\textwidth]{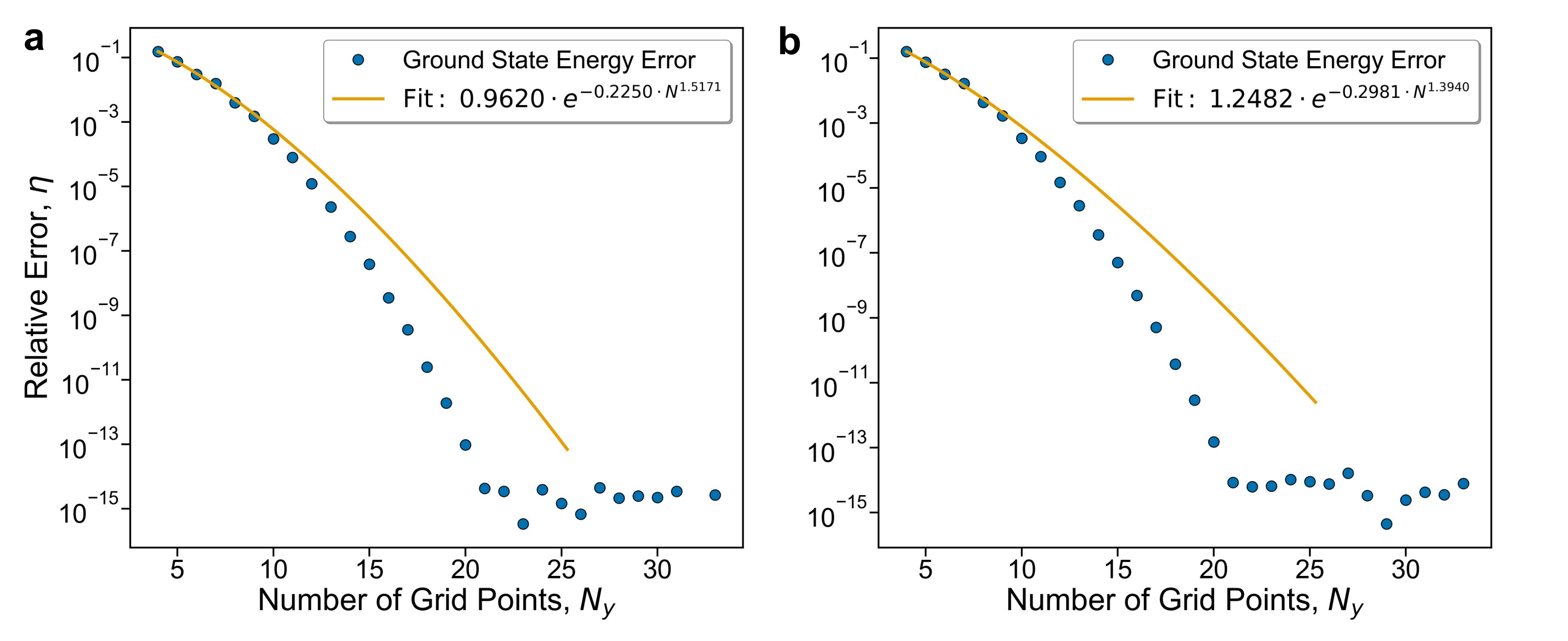}%
  \caption{
  Fitting analysis of the convergence rate with respect to the number of nuclear grid
points for $\omega_1$ = 3.0 and two different coupling strengths: (\textbf{a}) $g$ = 0.80 and (\textbf{b}) $g$ = 0.5. Shows a stretched exponential decay pattern for both coupling strengths, with a slightly faster decay rate for the weaker coupling strength ($g=0.5$).
  }	\label{fig:convergerate}
\end{figure*}

The convergence rates for model \RNum{1} with respect to the number of nuclear grid points (with $N_\text{s} = 8$) and with respect to the number of electronic states (fixing $N_y = 32$) are shown in \cref{fig:analytical1,fig:analytical2,fig:analytical3} for three different regimes, with corresponding APESs. The accuracy of the LDR method was quantified by the relative error, $ \eta = \lvert (E_{\text{LDR}} - E_{\text{analytical}})/(E_{\text{analytical}})\rvert $, where $\lvert \cdot \rvert $ denotes the absolute value. The strength of the non-Born-Oppenheimer effects is determined by the energy-scale separation $\omega_1$ and the coupling strength $g$.

Convergence with respect to the number of electronic states is highly sensitive to non-Born-Oppenheimer effects. Specifically, as the energy-scale separation increases from the comparable case ($\omega_1=1.0$) to the large-separation case ($\omega_1=10.0$), the error shows a clear exponential decay, with the number of electronic states required for convergence decreasing dramatically. Increasing the coupling strength $g$ requires more electronic states to reach the same level of accuracy. In the regime with large energy scale separation, a very high level of accuracy was also achieved for the excited states.
Convergence with respect to the nuclear grid is consistently achieved with a small number of points (approximately 20). A fitting analysis of the $\omega_1=3.0$ system (Fig.~\ref{fig:convergerate}) reveals a stretched exponential decay pattern. While a slightly faster decay rate is observed for weaker coupling ($g=0.5$), the difference is minor.

In summary, for the coupled harmonic oscillator model, the LDR method proves to be highly accurate and efficient, demonstrating robust performance even when ``electronic'' and ``nuclear'' energy scales are comparable. Its performance typically improves with larger energy scale separation. The efficiency is particularly evident in the small number of nuclear grid points.

\subsection{Nonlinearly Coupled Oscillator Models}\label{sec:nonlinear_models}

To benchmark the performance and stability of LDR in comparison to the traditional methods, we use two nonlinearly coupled harmonic oscillator models, $\hat{H}_\text{\RNum{2}}$ and $\hat{H}_\text{\RNum{3}}$, which feature asymmetric APESs. 
We consider six different methods 
\begin{enumerate}[label=(\arabic*), itemsep=0pt, parsep=0pt]
	\item Local Diabatic Representation with exact overlap matrix (LDR)
	\item LDR with approximate overlap matrix by the linked-product approximation (LDR+LPA)
	\item Born-Huang ansatz with nonadiabatic coupling only (NAC)
	\item Born-Huang ansatz with nonadiabatic coupling and diagonal Born-Oppenheimer correction (NAC + DBOC)
	\item Exact Born-Huang representation with NAC, DBOC, and second-order derivative coupling (NAC + DBOC + SDC)
	\item Crude Adiabatic Representation (CAR)
\end{enumerate}
For CAR, we select $y=0$ as the fixed reference ``nuclear'' geometry, meaning the adiabatic electronic states are defined as $\phi_{\alpha}(x;0)$. The performance of each method is quantified by the relative error of its first three eigenvalues, calculated as $\eta =\lvert (E_{\text{calc}} - E_{\text{exact}}) / E_{\text{exact}} \rvert$, where $E_{\text{exact}}$ represents the reference energies for both models, obtained from two-dimensional sine DVR calculations using 256 grid points per dimension in the range of $(-6,6)$.

\subsubsection{Model \RNum{2}}

Model \RNum{2} is described by the Hamiltonian with a nonlinear coupling:
\begin{figure}[htbp]
    \includegraphics[width=0.60\columnwidth]{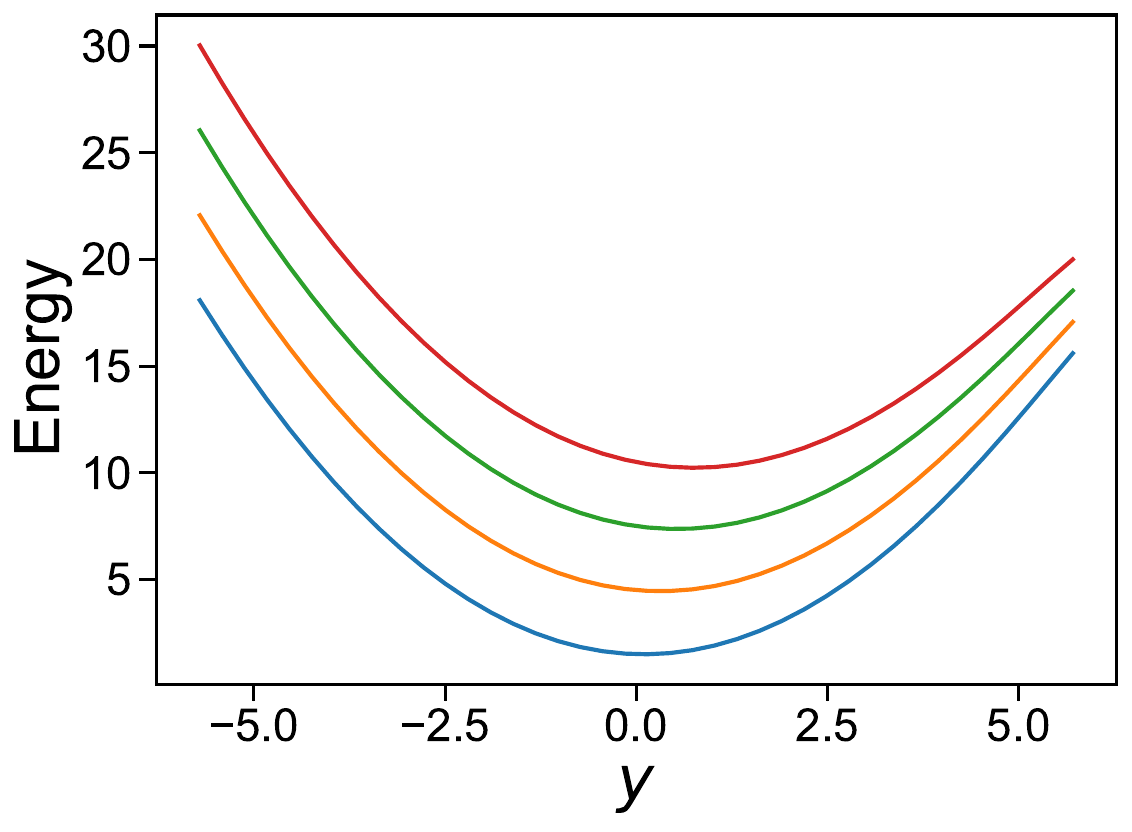}
\caption{APESs of model $\hat{H}_\text{\RNum{2}}$ for a representative parameter set: $\omega_1 = 3.0, g=0.5, \lambda=0.2$. The surface exhibits y-dependent curvature, non-uniform energy level spacing, and a potential minimum displaced from the origin. These features are shared across other parameter regimes.}
\label{fig:apes2}
\end{figure}
\begin{equation}
  \hat{H}_\text{\RNum{2}} = \frac{1}{2}\omega_1 (\hat{p}_x^2 + \hat{x}^2)+\frac{1}{2} (\hat{p}_y^2 + \hat{y}^2 )+ \frac{1}{2}g\hat{x}\hat{y} - \lambda \hat{y} \hat{x}^2
  \label{eq:H2}
\end{equation}
Although the eigenvalues of $\hat{H}_\text{\RNum{2}}$ are not analytically solvable, a key advantage is that its ``electronic'' eigenstates can still be determined analytically. This property allows the corresponding NAC elements to be also derived analytically. Similar to the model \RNum{1}, the adiabatic eigenstates are the displaced harmonic oscillator eigenstates with a ``nuclear-dependent'' displacement
\be 
\Delta(y)=-\frac{gy}{2\omega_1 - 4\lambda y}.
\ee
\begin{figure*}[htbp]
  \includegraphics[width=0.80\textwidth]{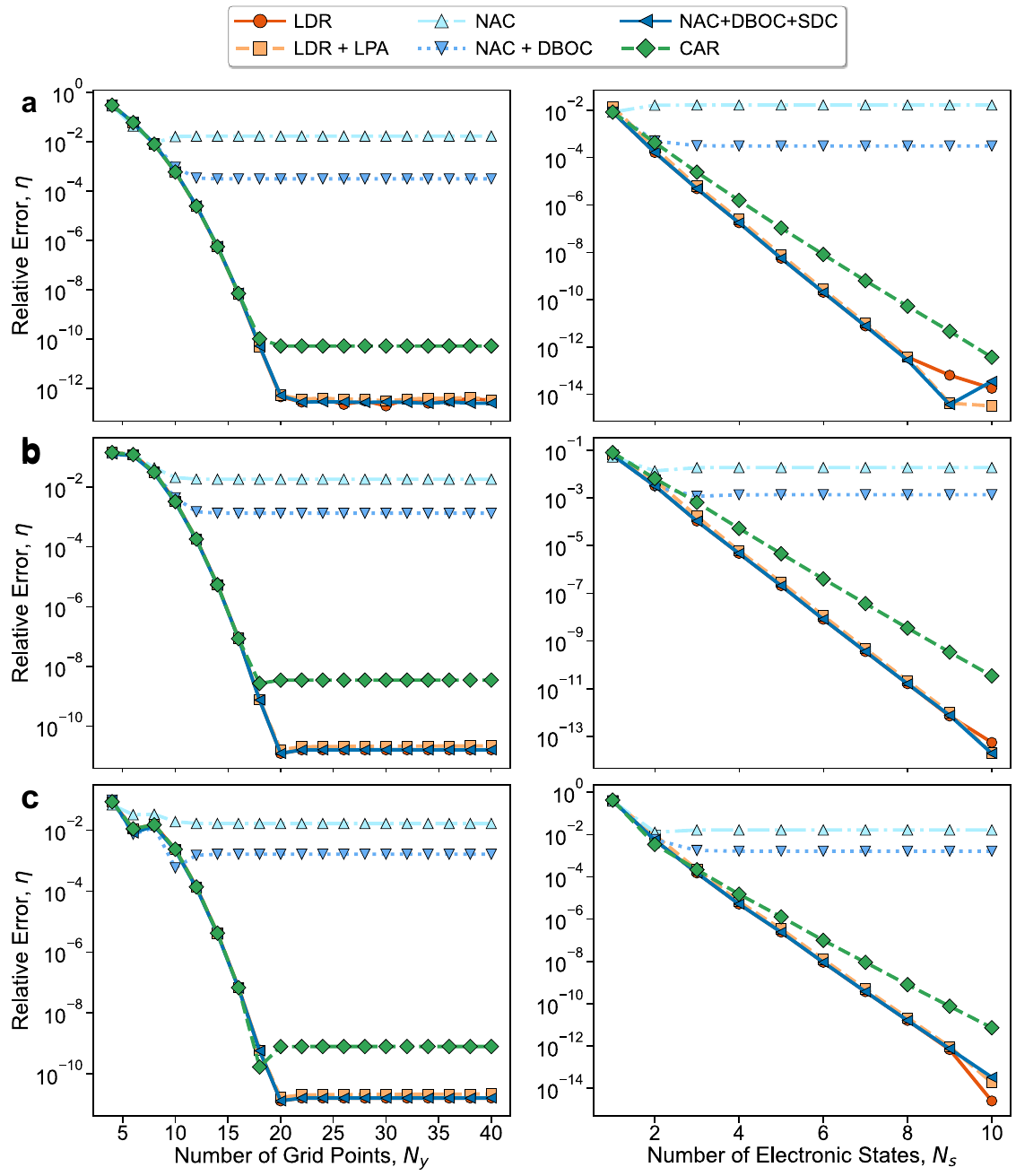}
\caption{Convergence of the LDR method for model $\hat{H}_\text{\RNum{2}}$ with the parameter set ($\omega_1 = 1.0, g=0.5, \lambda=0.05$). The panels show the results for the: (\textbf{a})~ground state, (\textbf{b})~first excited state, and (\textbf{c})~second excited state. The error in the ground state energy of LDR converges to $10^{-12}$ with approximately 20 nuclear grid points, and with 10 electronic states, it settles into the $10^{-13} \sim 10^{-14}$ error range.}
  \label{fig:convergence21}
\end{figure*}
\begin{figure*}[htbp]
  \includegraphics[width=0.80\textwidth]{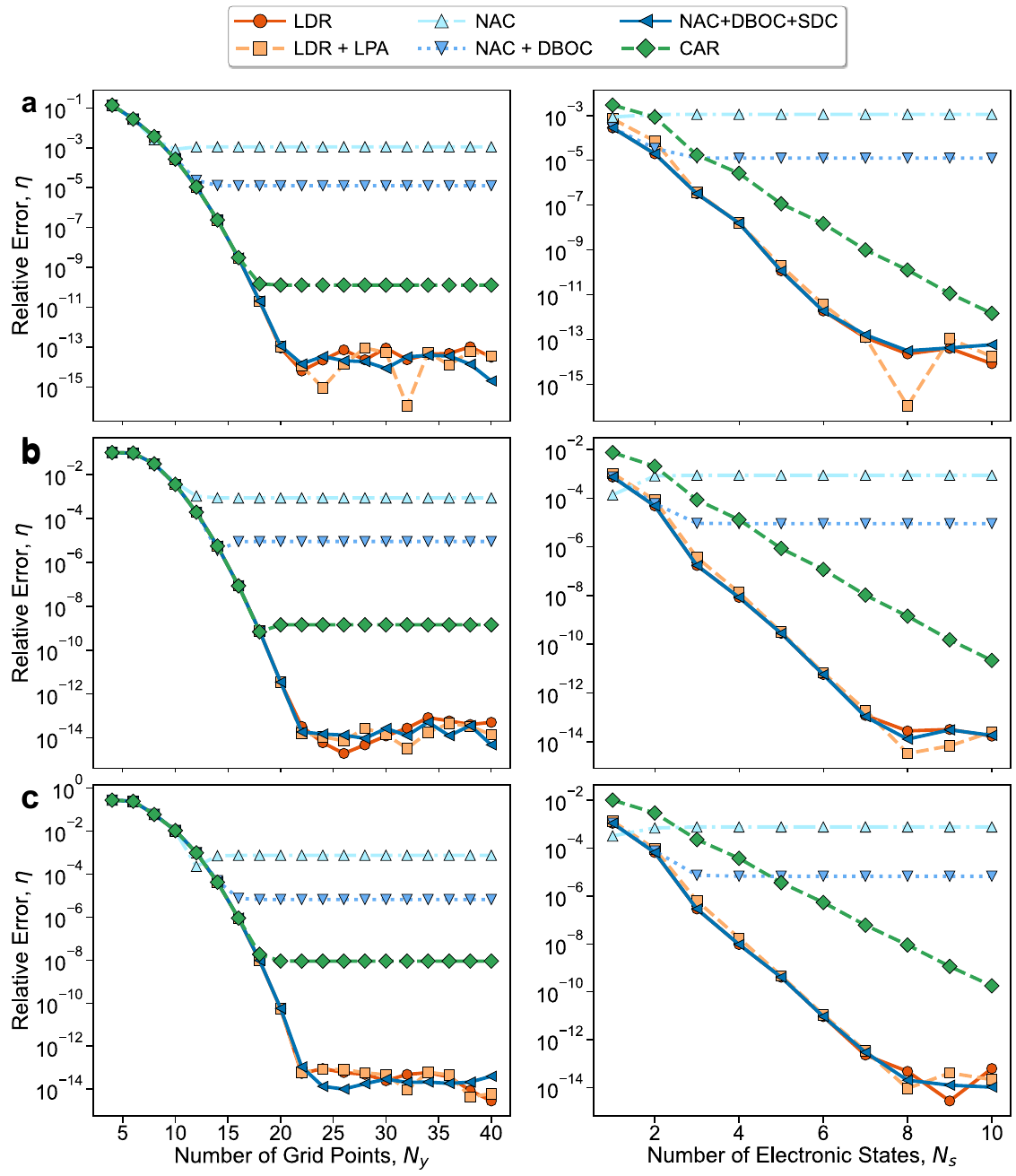}
  \caption{Convergence of LDR method for model $\hat{H}_\text{\RNum{2}}$ with parameter set ($\omega_1 = 3.0, g=0.5, \lambda=0.2$). The panels show the results for the: (\textbf{a})~ground state, (\textbf{b})~first excited state, and (\textbf{c})~second excited state. The error in the ground state energy of LDR converges to $10^{-14}$. This level of accuracy requires approximately 20 nuclear grid points and settles into the same error within 8 electronic states.}
  \label{fig:convergence22}
\end{figure*}
\begin{figure*}[htbp]
  \includegraphics[width=0.80\textwidth]{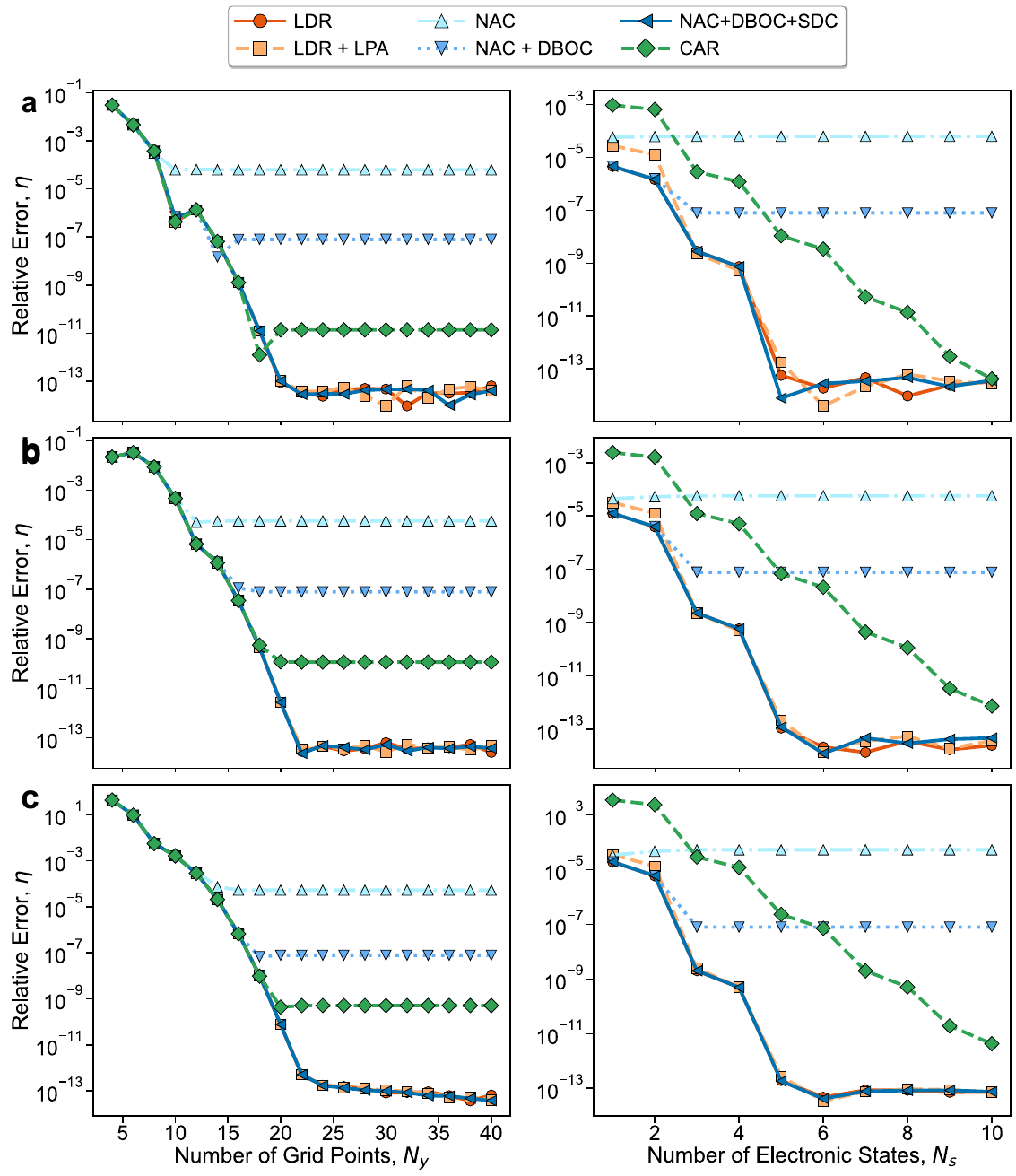}
\caption{Convergence of the LDR method for model $\hat{H}_\text{\RNum{2}}$ with the parameter set ($\omega_1 = 10.0, g=0.5, \lambda=0.5$). The panels show the results for the: (\textbf{a})~ground state, (\textbf{b})~first excited state, and (\textbf{c})~second excited state. The error in the ground state energy of LDR converges to $10^{-14}$. This level of accuracy requires approximately 20 nuclear grid points and settles into the same error within 5 electronic states.}
  \label{fig:convergence23}
\end{figure*}\noindent
The NAC matrix, $\mathbf{F}(y)$, is given by
\begin{equation}
    \mathbf{F}(y) = \Gamma_1(y) \left(\mathbf{a}-\mathbf{a}^{\text{T}}\right) + \Gamma_2(y) \left(\mathbf{a}^2-(\mathbf{a}^{\text{T}})^2\right),
    \label{eq:nac_h2}
\end{equation}
From the NAC matrix, the Diagonal Born-Oppenheimer Correction (DBOC) and the second-derivative coupling terms can be derived. This procedure yields the matrix $\mathbf{G}(y)$:
\begin{align}
    \mathbf{G}(y) ={}& \Gamma_1'(y) \left(\mathbf{a}-\mathbf{a}^{\text{T}} \right) + \Gamma_2'(y) \left( \mathbf{a}^2-(\mathbf{a}^{\text{T}})^2\right) \notag \\
    & + \Gamma_1(y)^2 \left( \mathbf{a}^2+(\mathbf{a}^{\text{T}})^2-\mathbf{a}^{\text{T}}\mathbf{a}-\mathbf{a}\mathbf{a}^{\text{T}} \right) \notag \\
    & + \Gamma_2(y)^2 \left( \mathbf{a}^4+(\mathbf{a}^{\text{T}})^4 -2(\mathbf{a}^{\text{T}}\mathbf{a})^2-\mathbf{a}^{\text{T}}\mathbf{a}-\mathbf{a}\mathbf{a}^{\text{T}} \right) \notag \\
    & + 2 \Gamma_1(y) \Gamma_2(y) \left( \mathbf{a}^3-\mathbf{a}(\mathbf{a}^{\text{T}})^2-\mathbf{a}^{\text{T}}\mathbf{a}^2+(\mathbf{a}^{\text{T}})^3\right).
    \label{eq:F_second_deriv}
\end{align}
In Eqs.~\eqref{eq:nac_h2} and \eqref{eq:F_second_deriv},  $\mathbf{a}$  and $\mathbf{a}^{\text{T}}$are the matrix representations of the  annihilation and creation  operators in the basis of the eigenstates $\ket{\phi_{\alpha}(y)}$, respectively. Their matrix elements are defined as $\mathbf{a}_{\beta,\alpha}=\sqrt{\alpha}\delta_{\beta,\alpha-1}$ and $\mathbf{a}^{\text{T}}_{\beta,\alpha}=\sqrt{\alpha+1}\delta_{\beta,\alpha+1}$, where integer indices $\alpha, \beta \in \{0, 1, 2, \dots\}$ represent the electronic states. And the coefficient $\Gamma_1(y) = -\frac{\Delta'(y)\,\xi(y)}{\sqrt{2}}$ represents the rate of change of the oscillator's displacement, $\Delta'(y)$, scaled by the function $\xi(y)$. The coefficient $\Gamma_2(y) = \frac{\xi'(y)}{2\xi(y)}$ represents the relative rate of change of the scaling function $\xi(y)$, where $\xi(y) = \left(1 - 2\lambda y/\omega_1 \right)^{1/4}$ and measures the parametric change of the $x$-mode frequency. Prime notation ($'$) denotes the derivative with respect to $y$. The coupling matrices are sparse: in the NAC matrix ($F_{\beta\alpha}$), couplings are restricted to nearest-neighbor ($\beta = \alpha \pm 1$) and next-nearest-neighbor ($\beta = \alpha \pm 2$) states; in the SDC matrix ($G_{\beta\alpha}$), which consists of diagonal DBOC and off-diagonal SDC elements, the couplings are more complex, connecting states separated by up to four quantum numbers (i.e., $\beta$ extends to $\alpha \pm 4$).

We use sine DVR for the ``nuclear'' coordinate, it is constructed from the ``particle-in-a-box'' sine eigenfunctions as a finite representation basis. Within this primitive basis set the matrix elements of the nuclear gradient operator, $\nabla_{\mu}$, has analytical form\cite{BECK2000multiconfiguration,LIGHT2000DiscreteVariable}
\be
\bigl(\nabla_{\mu}\bigr)_{kl}
= \vcenter{\hbox{$
  \begin{cases}
    \displaystyle\frac{4}{L}\,\frac{kl}{k^2 - l^2}\,, & k - l \text{ is odd}, \\
    0, & k - l \text{ is even}.
  \end{cases}$}}
\ee
The DVR basis set is constructed by a unitary transformation matrix, $\mathbf{U}$
\be
U_{mn} = \sqrt{\frac{2}{N + 1}} \sin\!\biggl(\frac{mn\pi}{N + 1}\biggr)
\ee
The indices $m , n\in \{0, 1, 2, \dots\ N-1\}$ where $N$ is the total number of grid points or basis functions used.
Applying this transformation yields the matrix elements of the nuclear gradient operator in DVR
\begin{equation}
\label{eq:grad-optimized}
\bigl(\nabla_{\mu}^{\mathrm{DVR}}\bigr)_{mn}
=
\sum_{k,l=0}^{N-1}
U_{mk}^{*}\,
\bigl(\nabla_{\mu}\bigr)_{kl}\,
U_{ln}
\end{equation}
With this nuclear momentum operator, all the non-Born-Oppenheimer terms $ 2F^{\mu}_{\beta\alpha}(\mathbf{R}) \nabla_{\mu} + G^{\mu}_{\beta\alpha}(\mathbf{R})$
in Eq.~\eqref{BH-T} , can be evaluated analytically.

Similar to the Eq.~\eqref{eq:overlap_full}, the electronic overlap matrix elements for the model $\hat{H}_\text{\RNum{2}}$ can be calculated analytically as\cite{ANSBACHER1959Note, CHANG2005new}
\begin{widetext}
\begin{subequations}\label{eq:ovlp_simplified}
\begin{align}
A_{0,0}(y,y')
 &= \sqrt{\frac{2r}{1+r^2}}
   \exp\!\Bigl[-\frac{r^2}{2(1+r^2)}\,\Delta_{y,y'}^2\Bigr]
 \label{eq:ovlp_simp_base}
 \\
A_{\beta,\alpha}(y,y')
 &=\frac{D_{y,y'}}{1+r^2}\,\sqrt{\frac{2}{\beta}}\,
   A_{\beta-1,\alpha}(y,y')
  + \frac{2r}{1+r^2}\,\sqrt{\frac{\alpha}{\beta}}\,
   A_{\beta-1,\alpha-1}(y,y')
  + \frac{r^2-1}{1+r^2}\,\sqrt{\frac{\beta-1}{\beta}}\,
   A_{\beta-2,\alpha}(y,y')
 \label{eq:ovlp_simp_rec1} 
 \\
 &= -\,\frac{r\,D_{y,y'}}{1+r^2}\,\sqrt{\frac{2}{\alpha}}\,
    A_{\beta, \alpha-1}(y,y')
  + \frac{2r}{1+r^2}\,\sqrt{\frac{\beta}{\alpha}}\,
    A_{\beta-1,\alpha-1}(y,y')
  - \frac{r^2-1}{1+r^2}\,\sqrt{\frac{\alpha-1}{\alpha}}\,
    A_{\beta,\alpha-2}(y,y')
 \label{eq:ovlp_simp_rec2}
\end{align}
\end{subequations}
\end{widetext}
Here, $r = \xi^2(y)/\xi^2(y')$, and $D_{y,y'}=\xi^2(y)\Delta_{y,y'}$. The integer indices $\alpha, \beta \in \{0, 1, 2, \dots\}$ represent the electronic states. The recursive calculation proceeds using relation Eq.~ \eqref{eq:ovlp_simp_rec1} for states where $\beta > 0$, and relation Eq.~ \eqref{eq:ovlp_simp_rec2} when $\beta = 0, \alpha > 0$. The recursion terminates at the base case $A_{0,0}$, with the boundary condition that any element with a negative index is defined to be zero.

We investigate three regimes based on the ratio between electronic and nuclear energy scales: where the electronic energy scale is significantly larger ($\omega_1=10.0, \lambda=0.5$), moderately larger ($\omega_1=3.0, \lambda=0.2$), and comparable ($\omega_1=1.0, \lambda=0.05$) to the ``nuclear'' scale. The coupling strength $g$ is fixed at $0.5$ in all cases. As a representative example, the APES for the moderately larger energy scale regime is shown in \cref{fig:apes2}. While the surfaces in other regimes are qualitatively similar, they differ primarily in the energy gap between the states, which becomes more pronounced as the electronic energy scale  ($\omega_1$) increases.
\begin{figure*}[!t]
  \includegraphics[width=0.65\textwidth]{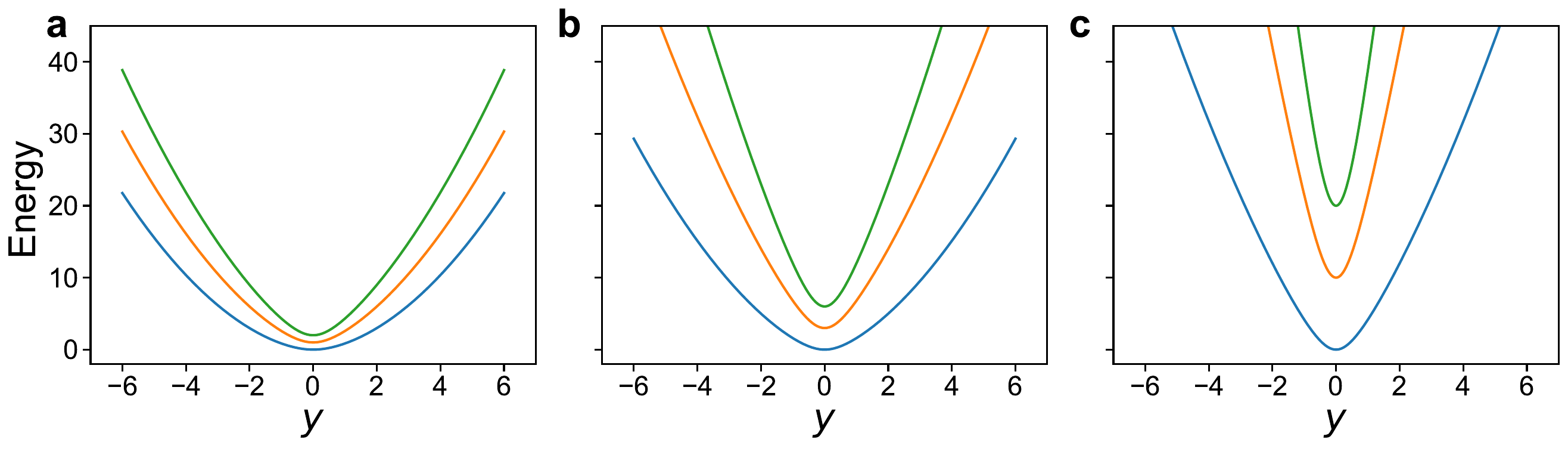}
  \caption{APESs of model $\hat{H}_\text{\RNum{3}}$ for three representative parameter sets: ({a})~$\omega_1 = 1.0, g=0.5, \lambda=1.0$; ({b})~$\omega_1 = 3.0, g=0.5, \lambda=3.0$; and ({c})~$\omega_1 = 10.0, g=0.5, \lambda=10.0$. The surfaces for this model are symmetric about $y=0$ and exhibit steep curvatures that change significantly as a function of $y$. All three panels share the same y-axis scale.}
  \label{fig:apes3}
\end{figure*}

The convergence rate for model \RNum{2} for the number of nuclear grid points (with $N_\text{s} = 8$ electronic states) and for the number of electronic states (fixing $N_y = 32$) are shown in \cref{fig:convergence21,fig:convergence22,fig:convergence23} for three different regimes.
Overall, the LDR and LDR+LPA methods achieve an accuracy comparable to the exact Born-Huang representation, and the convergence behavior of these three methods is similar. For all three approaches, convergence with respect to the nuclear grid is consistently achieved with approximately 20 points, irrespective of the specific regime. The primary distinction between the regimes, therefore, lies in the convergence with respect to the number of electronic states, which varies significantly depending on energy scales.

In the regime where energy scales are comparable ($\omega_1=1.0, \lambda=0.05$), both the LDR and the exact Born-Huang representation achieves a ground-state accuracy on the order of $10^{-13}$ using 8 electronic states. And 10 for the first and second excited states. As the energy scale separation increases ($\omega_1=3.0, \lambda=0.2$ and $\omega_1=10.0, \lambda=0.5$), accuracy for the ground state improves to the order of $10^{-14}$ with fewer electronic states.
Both LDR and LDR+LPA exhibit a similar exponential convergence behavior to that observed in the coupled harmonic oscillator model.
Comparison with NAC and NAC + DBOC reveals the importance of SDC when applying Born-Huang representation for vibrational eigenvalue problems. The NAC and NAC+DBOC methods are insufficient to capture non-Born-Oppenheimer effects in the regime with comparable energy scales, yielding large errors of $10^{-2}$ and $10^{-3}$, respectively. As the energy scale separation increases, their performance improves slightly. This trend aligns with physical intuition: a larger separation signifies weaker electron-nuclear coupling.
CAR initially shows a similar convergence rate as LDR and exact Born-Huang representation with respect to the number of nuclear grid points, but does not reach the same level of accuracy in the plateau. Moreover, it shows a slow convergence rate with respect to the number of electronic states.

\subsubsection{Model \RNum{3}}

The third model is described by the Hamiltonian

\begin{equation}
  \hat{H}_\text{\RNum{3}} = \frac{1}{2}\omega_1 (\hat{p}_x^2 +\hat{x}^2 )+ \frac{1}{2} (\hat{p}_y^2 +\hat{y}^2 )+ \frac{1}{2}g\hat{x}\hat{y} + \lambda \hat{x}^2 \hat{y}^2 \,.
  \label{eq:H3}
\end{equation}
It contains a different nonlinear mode-coupling, nevertheless, the adiabatic states are still displaced harmonic oscillator states, 
with the displacement function given by 
\be \Delta(y)=-\frac{gy}{2\omega_1 + 4\lambda y^2}
\ee and the scaling function 
\be 
\xi(y)=\left(1 + \frac{2\lambda y^2}{\omega_1} \right)^{\frac{1}{4}}
\ee 
 Consequently, the nonadiabatic couplings and the electronic overlap matrix can be simply obtained from Eqs.~\eqref{eq:nac_h2}–\eqref{eq:ovlp_simplified}.
 % by replacing the previous model's functions, $\Delta_2(y)$ and $\xi_2(y)$, with $\Delta_3(y)$ and the new term 
\begin{figure*}[htbp]
  \includegraphics[width=0.80\textwidth]{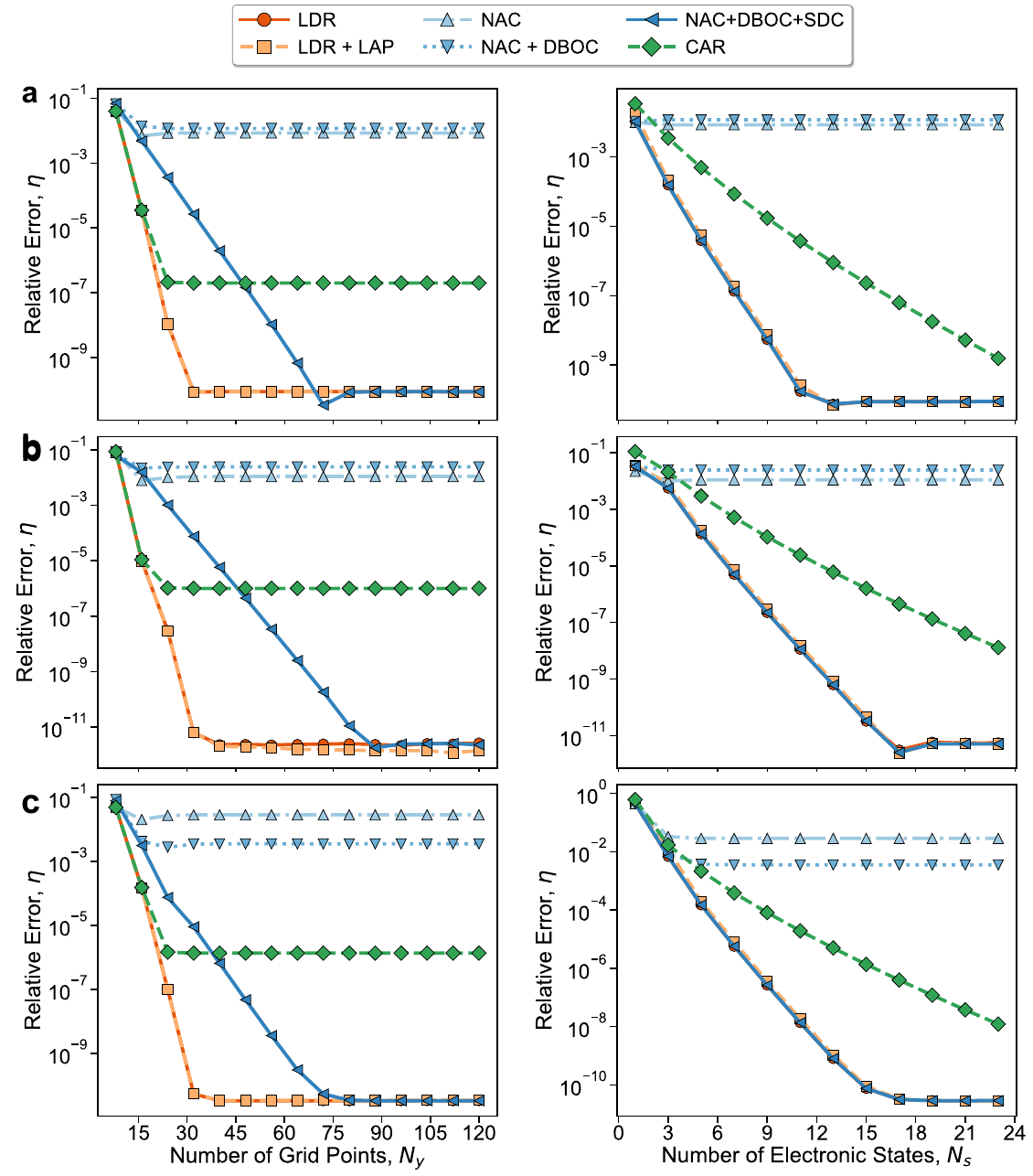}
\caption{Convergence of the LDR method for Model \RNum{3} with parameters $\omega_1 = 1.0$, $g=0.5$, and $\lambda=1.0$. The panels show results for the (\textbf{a}) ground state, (\textbf{b}) first excited state, and (\textbf{c}) second excited state. For the LDR method, the ground state error converges to a relative error of $10^{-11}$. This level of accuracy requires approximately 30 nuclear grid points and achieved with 13 electronic states.}
  \label{fig:convergence31}
\end{figure*}
\begin{figure*}[htbp]
  %\centering
  \includegraphics[width=0.80\textwidth]{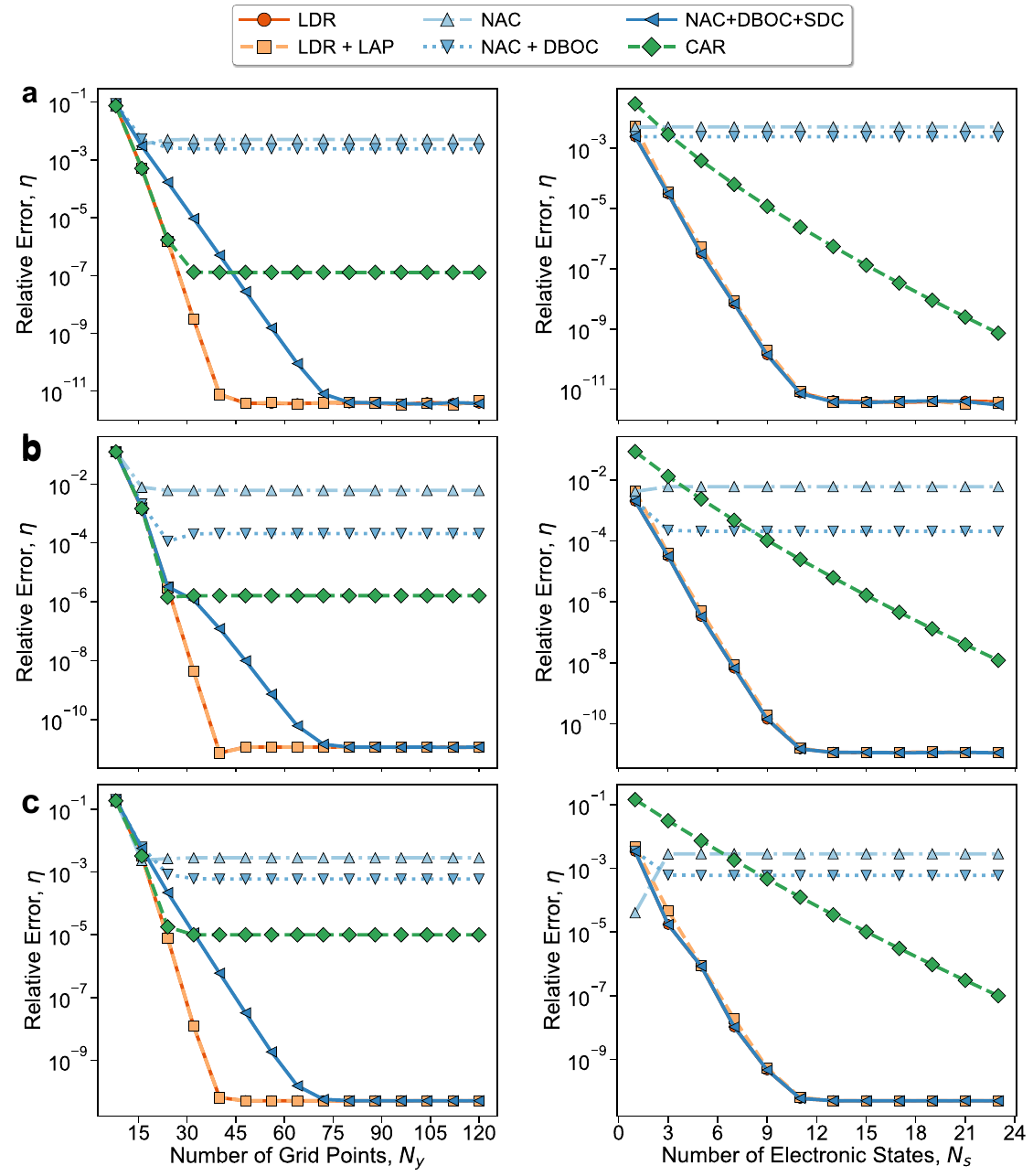}
  \caption{Convergence of LDR method for model $\hat{H}_\text{\RNum{3}}$ with parameter set ($\omega_1 = 3.0, g=0.5, \lambda=3.0$). The panels show the results for the: (\textbf{a})~ground state, (\textbf{b})~first excited state, and (\textbf{c})~second excited state. For LDR method, the ground state error converges to a relative error of $10^{-13}$. This level of accuracy requires approximately 40 nuclear grid points and achieved with 13 electronic states.}
  \label{fig:convergence32}
\end{figure*}
\begin{figure*}[htbp]
  %\centering
  \includegraphics[width=0.80\textwidth]{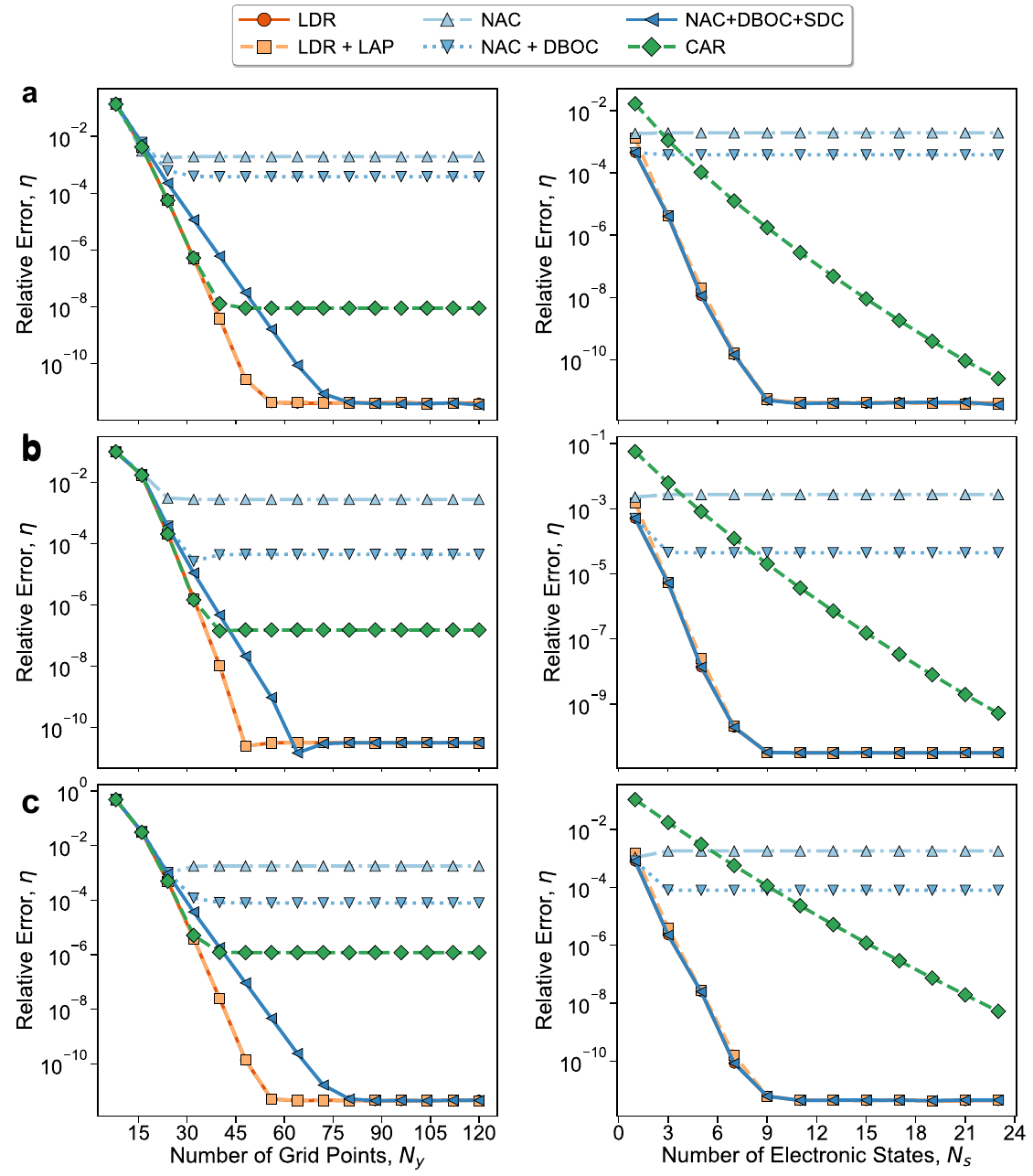}
  \caption{Convergence of LDR method for model $\hat{H}_\text{\RNum{3}}$ with parameter set ($\omega_1 = 10.0, g=0.5, \lambda=10.0$). The panels show the results for the: (\textbf{a})~ground state, (\textbf{b})~first excited state, and (\textbf{c})~second excited state. For LDR method, the ground state error converges to a relative error of $10^{-12}$. This level of accuracy requires approximately 50 nuclear grid points and achieved with 9 electronic states.}
  \label{fig:convergence33}
\end{figure*}

%For model $\hat{H}_\text{\RNum{3}}$, 
Similar as in models \RNum{1} and \RNum{2}, we investigate three regimes where the energy scale of the ``electrons'' is larger ($\omega_1=10.0, \lambda=10.0$), moderately larger ($\omega_1=3.0, \lambda=3.0$) , and comparable ($\omega_1=1.0, \lambda=1.0$) to the ``nuclei''. 
The coupling strength $g$ is fixed at $0.5$. The corresponding APESs are shown in \cref{fig:apes3}.
Convergence rate for model \RNum{3} for the number of nuclear grid points (with $N_\text{s} = 16$ ) and for the number of electronic states (fixing $N_y = 90$) are shown in \cref{fig:convergence31,fig:convergence32,fig:convergence33}. Compared to model \RNum{2}, the non-Born-Oppenheimer effects are stronger in this model as it requires much more electronic states to converge, even for the ground state energy.

%in terms of both convergence limits and rates.

We found that LDR-based methods converge significantly faster even compared with the exact Born-Huang representation with respect to the number of nuclear grid points across all parameter sets, both methods converge to the same limit in the complete nuclear basis set limit as expected. Their convergence rates with respect to the number of electronic states, remain comparable.
The LPA is surprisingly accurate across all parameter regimes, introducing only a negligible error that does not impact overall performance.
The convergence behavior with respect to the nuclear grid clearly highlights the efficiency of the LDR method over the Born-Huang approach across three regimes. This is most evident in the comparable energy separation regime ($\omega_1 = 1.0, \lambda=1.0$). With 30 nuclear grid points, a typical number for DVR basis set, the error of LDR is about six orders of magnitude smaller than that of the Born-Huang approach, which requires approximately three times more grid points to reach a similar accuracy. Note that in ab initio modeling, increasing the number of grid points not only increases the matrix size but also the number of quantum chemistry computations.
This difference in convergence rates can be explained by the different ways in which LDR and Born-Huang describe nonadiabatic effects. The non-Born-Oppenheimer effects are described by the electronic overlap matrix in LDR and by the nonadiabatic coupling terms (NAC, DBOC, and SDC) in the Born-Huang approach. The nonadiabatic coupling terms exhibit sharper features and more rapid variations than the overlap matrix, thus requiring a finer spatial grid. As the energy scale separation increases, the nonadiabatic coupling terms becomes smoother leading to a better convergence for the Born-Huang method. This feature hints the challenges of the Born-Huang approach to describe conical intersections, where the energy gap vanishes.

The performance of the NAC and NAC+DBOC methods is overall similar to that in Model $\hat{H}_{\text{\RNum{2}}}$, but with a few notable differences. Interestingly, for the ground-state error, the NAC+DBOC method performs worse than the NAC method alone in the regime with comparable energy scales. 
Across all regimes, the performance of the CAR method in this model is largely similar to its behavior in Model $\hat{H}_{\text{\RNum{2}}}$. In the strong-coupling regime, although its error appears to decrease as more electronic states are included, the convergence rate is much slower than that of the LDR and exact Born-Huang representation, as it achieves an accuracy of only $10^{-9}$ even with 23 electronic states. The performance of the CAR method improves with the energy-scale separation. In the regime with the largest energy scale separation, it achieves an accuracy of $10^{-11}$ with 23 electronic states; this result is only one to two orders of magnitude less accurate than that of the LDR method.

In summary, the benchmark results on nonlinear coupled oscillator models suggests with the following observed trend in accuracy:
\begin{align*}
  \text{LDR} &\approx \text{LDR+LPA} \approx \text{NAC+DBOC+SDC} \\
  &> \text{CAR} \gg \text{NAC+DBOC} > \text{NAC}
\end{align*}
In terms of convergence rate, LDR also exhibits a significant efficiency advantage in all systems . The general trend in convergence speed can be summarized as follows:
\begin{align*}
 & \text{LDR} \approx \text{LDR+LPA} \gtrsim \text{NAC+DBOC+SDC}\gg \text{CAR}
\end{align*}

\section{Conclusion}\label{sec:conclusion}

We have performed a systematic benchmark of the Local Diabatic Representation (LDR) method comparing its performance to the conventional methods based on the Born-Huang ansatz as well as the crude adiabatic representation. Our results demonstrate that LDR is a robust, efficient, and highly accurate method for solving nonadiabatic eigenvalue problems.

With strong derivative couplings, LDR consistently requires fewer grid points and converges faster than the Born-Huang ansatz.
Its practical advantages include the exponential convergence with respect to both the number of electronic states and the number of grid points and avoiding the singular derivative couplings, and can be used for any gauge fixings —make it particularly useful for tackling challenging nonadiabatic systems, e.g., molecules involving conical intersections.

\begin{acknowledgements}
We thank Dr. Vitaly Rassolov for suggesting the benchmark studies. This work is supported by the National Natural Science Foundation of China (Grant No. 22473090 and 92356310).

\end{acknowledgements}

\appendix
\bibliography{benchmark,dynamics}
\end{document}